\documentclass[prb,preprint,longbibliography]{revtex4-1} 
\usepackage[usenames,dvipsnames]{color}
\usepackage{amsmath} 

\usepackage{graphicx} 

\begin{document}

\title{Polarized nuclear target based on parahydrogen induced polarization}

\author{D.~Budker}
 \email{budker@berkeley.edu}
 \affiliation{Department of Physics, University of California at
Berkeley, Berkeley, California 94720-7300}
 \affiliation{Nuclear Science Division, Lawrence
 Berkeley National Laboratory, Berkeley, California 94720}

\author{M.~P.~Ledbetter}
\affiliation{Department of Physics, University of California at Berkeley, Berkeley, California 94720-7300}

\author{S.~Appelt}
\affiliation{Central Institute for Electronics, Research Center J$\ddot{u}$lich, D-52425 J$\ddot{u}$lich, Germany}

\author{L.~S.~Bouchard}
\affiliation{Department of Chemistry and Biochemistry, California NanoSystems
Institute, Biomedical Engineering IDP, Jonsson Comprehensive Cancer Center,
University of California, Los Angeles, CA 90095, USA}

\author{B.~Wojtsekhowski}
\affiliation{Thomas Jefferson National Accelerator Facility, Newport News, VA 23606}

\date{\today}

\begin{abstract}
We discuss a novel concept of a polarized nuclear target for accelerator fixed-target scattering experiments, which is based on parahydrogen induced polarization (PHIP). One may be able to reach a 33\% free-proton polarization in the ethane molecule.
The potential advantages of such a target include operation at zero magnetic field,
fast ($\sim$100 Hz) polarization reversal, and operation with large intensity of an electron beam.
\end{abstract}

\maketitle

\section{Introduction}
\subsection{Motivation}
Polarized nuclear targets, recently reviewed in Refs.~[\cite{Crabb:1997cy, Goertz:2002vv}], play an important role
in particle physics in investigations of fundamental symmetries and dynamical aspects of known interactions,
in particular, those aimed at studying the physics of the nucleon structure functions \cite{PDG}.
The structure functions $g_1$ and $g_2$ for both proton and neutron are important for understanding of the quantum chromodynamical (QCD) nature of the nucleon spin.
Currently, a number of accelerator based scattering experiments explore polarization observables for the structure-function investigations \cite{Ans95,PDG}.
A new way to study the nucleon spin emerged in the last decade: the SIDIS (semi-inclusive deep inelastic scattering with transversely polarized nucleons) experiments are probing the effect of nucleon polarization on the azimuthal variation of the leading meson-production cross section in deep inelastic electron-nucleon scattering.
This effect will allow measurement of the quark orbital angular moment (OAM), as discussed by Sivers \cite{Siv90} and Collins \cite{Col2002}.
A nonzero value of OAM means that quarks move coherently, at least a fraction of the time, around the spin direction.
The concept of OAM is attractive,  however, reliable SIDIS data are needed to validate it.
While important measurements have been performed at the Deutsches Elektronen Synchrotron (DESY) \cite{HERMES-05,HERMES-09}
and the European Organization for Nuclear Research (CERN) \cite{COMPASS2005,COMPASS2010}, the experimental accuracy
needs to be improved by orders of magnitude.
The Thomas Jefferson National Accelerator Facility (JLab) at Newport News, Virginia has a large program of SIDIS experiments with a 12 GeV electron beam \cite{JLab12GeV}.
The hadron facility FAIR at Darmstadt, Germany has a physics program
for studies of the antiproton interaction with polarized proton and neutron targets \cite{FAIR}.

As far as weak interactions are concerned, a parity non-conserving term in deep inelastic electron-scattering (DIS) cross section appears for a polarized electron beam due to the interference of electromagnetic and weak-neutral-current processes ($\gamma -Z$ interference). It was first observed at the Stanford Linear Accelerator Center (SLAC) with an unpolarized target (see Ref.\ \cite{Cli97} and references therein).  Complementary parity-violating asymmetries arise for an unpolarized beam and a longitudinally polarized target, as predicted in Ref.\ \cite{ZEUS95}. However, these are hard to measure because the event rates in polarized-target experiments are usually low. This effect potentially provides access to the parity violating structure functions \cite{Ans95}, which were measured in deep inelastic neutrino/antineutrino-nucleon scattering and high momentum transfer electron-proton DIS \cite{Bil75}.
It will be interesting to verify these using an unpolarized electron beam and a polarized nucleon target.

\subsection{Polarized nuclear targets}
The key parameters of a nuclear target are:
the areal density of the target material ($N$, in nucleons/cm$^2$), the usable beam intensity ($I$, in electrons/s), the electron-nucleon luminosity, which is the product of these quantities
\begin{equation}
L=N\times I,
\end{equation}
usually given in Hz/cm$^2$; the average degree of nucleon polarization ($P$);
and, finally, the impact of the target magnet on particle detection, including restrictions on the detector solid angle
and deflection of the particle trajectories by the target's magnetic field.
The average nucleon polarization is
\begin{equation}
P=P_N\times D,
\end{equation}
where $P_N$ is the polarization degree of the polarizable nucleons,
and $D$ is a dilution factor, which takes into account the fraction of polarizable nucleons in the target material.

There are several experiments with polarized targets specifically designed to study the DIS/SIDIS processes:
\begin{enumerate}
  \item Internal polarized-target experiment HERMES at the HERA storage ring at DESY
operated at a low luminosity of $L\approx 1\times 10^{32}\ $Hz/cm$^2$.
Both polarized-proton~\cite{HERMES-H-D-targets} and polarized-neutron targets~\cite{HERMES-He3-target} were developed.
A polarization of protons $P_p\approx 0.9$ was achieved with a dilution factor $D=1$.
These targets allow an almost $4\pi$ solid angle for the detectors and require only a modest magnetic field of 3~kG
on the target in spite of the high intensity of the electron beam.

  \item High-energy, low-intensity muon-beam polarized-target experiments~\cite{SMC-experiment, COMPASS-experiment},
which provide a modest luminosity of $L \approx 5\times 10^{32}\ $Hz/cm$^2$ (achieved with the beam intensity of $3\times 10^7$ muons per second, target length of almost 1.5 m and total areal density of $\sim$50 g/cm$^2$), $P_p\approx 0.9$,
and a dilution factor of $D\approx 0.16$.
This lower dilution factor reflects a large fraction of unpolarized nucleons in that target.
The magnetic field on the target could be as low as 5~kG during data taking.
Because of the low beam intensity, the target polarization lasts for hundreds of hours.
Re-polarization of the target is performed in a 50~kG field, requires several hours, and is usually
repeated every two days. A combination of the high beam energy and a modest value of the target's magnetic field
allows for the required detector acceptance.

  \item Polarized-target experiments with up to 32~GeV electron beams at the Stanford Linear Accelerator Center
(SLAC)~\cite{E143-experiment, E154-experiment, E155-experiment},
which used a $^3$He target \cite{E142-target}
with $P_n \approx 0.3$, effective dilution $D \approx 0.125$, $L \approx 3\times 10^{36}\ $Hz/cm$^2$,
and NH$_3$/LiD targets~\cite{E143-target} with $P_p \approx 0.9$, $D\approx 0.16$, and
$L \approx 7 \times 10^{35}\ $Hz/cm$^2$ for studies of inclusive deep inelastic scattering.

  \item There are several polarized targets for experiments with the high-intensity medium-energy electron beam
at JLab \cite{CEBAF-accelerator}: \\
\hskip 1in i) An advanced version of the SLAC $^3$He target with neutron polarization
of $P_n \approx 0.43$ and $D\approx 0.25$.
Beam intensities up to 15 $\mu$A were used resulting in $L \approx 3\times 10^{36}\ $Hz/cm$^2$.
The latest development with circulating-gas approach allows even higher luminosity \cite{ConvectionTarget}.\\
\hskip 1in ii) A polarized NH$_3$/ND$_3$ target from the SLAC experiments and a similar one
for the large-acceptance detector CLAS \cite{CLAS-NH3}. \\
\hskip 1in iii) A ``frozen-spin'' target for the CLAS detector \cite{CLAS-FrozenSpin}. \\
\hskip 1in iv) A novel HD polarized target for the CLAS detector.
This target is designed to provide a high degree of nucleon polarization, a low dilution factor of 0.33, and
a modest magnetic field on the target of 10~kG.
The target\cite{CLAS-HD-proposal} is projected to operate at a luminosity of $L\approx 1\times 10^{34}\ $Hz/cm$^2$.
\end{enumerate}

The JLab accelerator will be upgraded to 12 GeV.
A large set of polarized-target experiments (both DIS and SIDIS) is already approved, and some experiments
are already in preparation~\cite{PAC38}.
In addition to the upgrade of the polarized targets, there is a program developing a set of new detectors.
It is apparent that, thanks to the advances in the $^3$He target, investigation of the neutron structure has a good outlook.
Higher-luminosity $^3$He targets additionally allow operation in small magnetic fields (typically, 25 G).

\subsection{Polarization and magnetic-field issues}
Investigation of the nucleon spin structure requires the study of both inclusive and semi-inclusive processes,
which means that average nucleon polarization is important, and that unpolarized nuclei in the target cause
a loss of experimental accuracy.
For example, in a polarized NH$_3$ target, the average nucleon polarization is only 1/6 of the free-proton polarization.
All known targets with polarized protons and some with polarized neutrons operate at low temperatures,
where the heat load from the incident particle beam becomes an issue, in addition to the radiation damage in the target material.

Magnetic fields required in proton polarized targets should be large, 50 kG, with a $10^{-4}$ uniformity.
Such requirements result in limited acceptance usable for the experiments' detectors.
A reduction of the field is possible for a low-luminosity regime.
In many currently running and planned experiments, nuclear spins in the target should be oriented transversely to the beam direction.
A transverse magnetic field bends the beam trajectory and leads to an additional background in the detector.
Even more importantly, the field deflects the secondary particles of interest, whose energies are much lower than
the incident-beam energy.
This problem is significant for medium-energy experiments at JLab, where the secondary particles have energies in
the 2-4 GeV range.
With a transversely polarized NH$_3$ target, the deflection of a 2 GeV pion is about 15 degrees, which is too large
for a SIDIS experiment.

\subsection{Basic parameters of a target and luminosity}
Productivity of an experiment depends on the target's Figure-of-Merit ($FoM$), which
is defined by luminosity, target polarization, and the dilution factor as
\begin{equation}
FoM = L \times P^2 = L \times (P_N\times D)^2.
\end{equation}
A minimum $FoM$ required for a productive future SIDIS experiment,
even with a large solid angle detector as CLAS, is about $10^{33}$ Hz/cm$^2$.
With the two protons in a C$_2$H$_4$ molecule polarizable by the parahydrogen technique discussed below, a projected dilution factor is 0.07.
As a result, the above $FoM$ translates into a required luminosity of $2 \times 10^{36}$ Hz/cm$^2$.
If beam induced target warm-up and depolarization are not important, the beam intensity at JLab
could be up to $100\ \mu$A. With such a large beam intensity, the required amount of material is small, 3~mg/cm$^2$.
Assuming a target area of 1.5 cm$^2$, the total mass of the material in the target is 5~mg.
Using a conservative value for the spin-relaxation time of $T_1\approx\ $10 ms, one can see that the reactor
should produce at least 500~mg/s of polarized C$_2$H$_4$ material.
In this scheme, the power deposition in the target by the beam is $\approx 1\ $W, and
the temperature rise is small due to a significant mass flow.

An ideal target for polarized-target experiments would (i) consist solely of protons (a simple nucleus) at high density and with (ii) full polarization. Moreover, it would allow for (iii) fast and efficient reversal of polarization (necessary to control systematic effects), (IV) would not require cryogenic temperatures to operate, and (V) would not require tesla-size magnetic fields hindering one's ability to control the polarization.

Here, we propose a concept of a polarized target that potentially addresses all of these requirements, with the exception of (i). The idea is based on the concept of parahydrogen induced polarization (PHIP) of nuclei\cite{Bowers86,Nat97}, and in particular, its zero-field version\cite{The2011}.

\section{The concept of the target}
\label{Sec:Concept}

\begin{figure}[h!]
   \begin{center}
    \includegraphics[height=4.5 in]{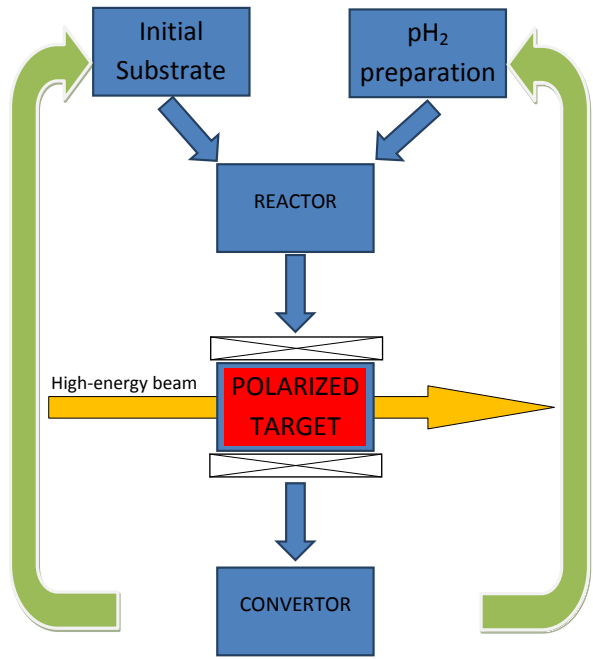}
    \caption{A block diagram of the proposed arrangement.}
    \label{Fig:Scheme}
    \end{center}
\end{figure}
The basic idea of the proposed target is illustrated in Fig.\ \ref{Fig:Scheme}. Pure parahydrogen (pH$_2$) is mixed with the initial substrate, for example, compressed $^{13}$C-enriched acetylene (C$_2$H$_2$) gas or liquid, pre-loaded with an appropriate catalyst (Sec.\ \ref{Sec:hydrogenation},\ref{Sec:Optimization}). The mixing occurs in the Reactor, where a catalyst-assisted hydrogenation reaction takes place, adding the two hydrogen atoms across the carbon-carbon double bond.

Upon hydrogenation, the two protons coming from a parahydrogen molecule (and which are now incorporated in the product, for example, C$_2$H$_4$ molecule) are initially in the singlet state, however, the combined spin state of $^{13}$C and the protons undergoes evolution under the scalar ($J$-coupling) Hamiltonian. This evolution results in a spin state that is devoid of net dipole polarization; however, such polarization can be induced \cite{Led2009} by applying an appropriate magnetic pulse (a typical pulse would be of DC magnetic induction $\sim$1 G and a millisecond duration).


Once the hydrogenation reaction is complete, the product flows into the actual target volume. Once there, it is exposed to a DC pulse induced by the magnetic coil (shown in Fig.\ \ref{Fig:Scheme} surrounding the Polarized Target). After the DC pulse, proton polarization will oscillate with a characteristic $J-$coupling frequency on the order of 100 Hz, while undergoing relaxation with a typical decay time of several seconds. The direction of the polarization is collinear with the direction of the magnetic-field induction in the pulse.  The polarization oscillation represents the fast reversal of the polarization in the target. Once the polarization oscillation decays, the contents of the target is emptied into the converter, while a fresh batch of hydrogenated material flows into the target volume. The used product is chemically regenerated (dehydrogenated) into substrate and hydrogen in the Converter, and can be reused for subsequent cycles.


\section{$J$-coupling induced polarization oscillations}
\begin{figure}[h!]
   \begin{center}
    \includegraphics[height=4 in]{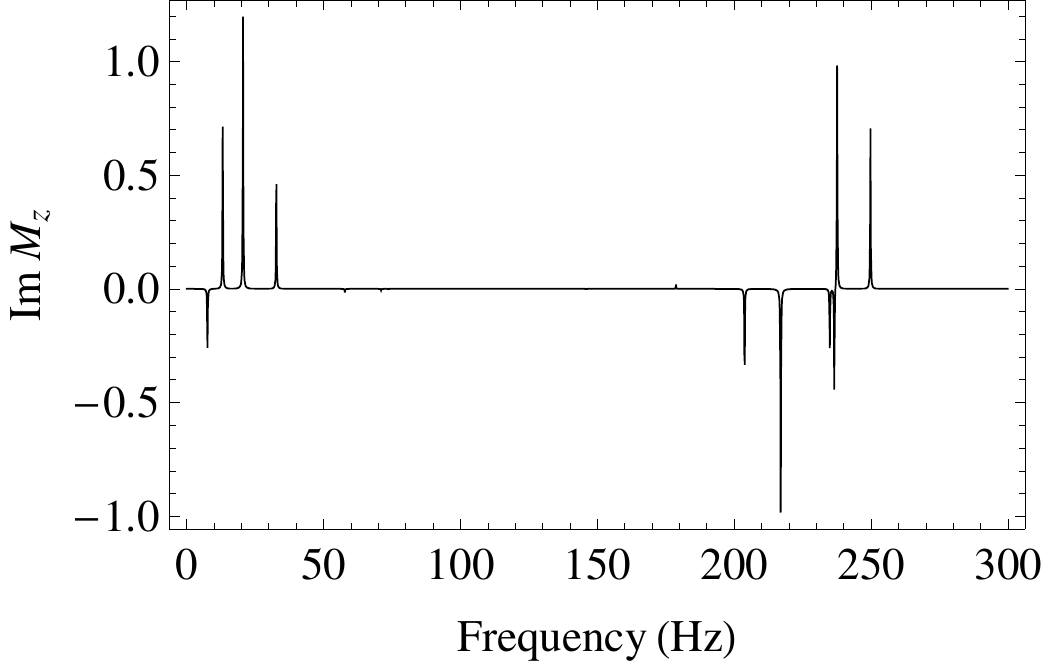}
    \caption{Imaginary component of the zero-field PHIP enhanced spectrum of ethylene, synthesized from singly $^{13}$C labeled acetylene via addition of parahydrogen.  This simulation shows oscillations of the observable nuclear magnetization at several frequencies corresponding to relative reversal of proton and $^{13}$C spins.}
    \label{Fig:Ethene_PHIP_simulation}
    \end{center}
\end{figure}

As mentioned above, a promising candidate for a target polarized via parahydrogen is ethylene (C$_2$H$_4$), synthesized from $^{13}$C labeled acetylene ($^{13}$C$^{12}$CH$_2$) via addition of parahydrogen.  Figure  \ref{Fig:Ethene_PHIP_simulation} shows the zero-field NMR spectrum of such a compound, following application of a DC pulse of magnetic field.  After such a pulse, magnetization oscillates along the direction of the pulse field at several frequencies, corresponding to rapid reversal of $^1$H and $^{13}$C spins with respect to each other.  The different frequencies displayed here should also show up in a Fourier transform of the scattered signal.  The zero-field NMR signal could thus be used as a diagnostic/reference in scattering experiments. We note that stray magnetic fields, including the magnetic field of the high-energy beam itself may modify the spectrum, so it will be important to control the fields at a milligauss level.

\section{Hydrogenation and dehydrogenation reactions}
\label{Sec:hydrogenation}
Parahydrogen-induced polarization (PHIP) proceeds by hydrogenation across an unsaturated carbon-carbon bond.  For example, propene (C$_3$H$_6$) can be hydrogenated to yield propane (C$_3$H$_8$) using a suitable catalyst. Recently, it has been demonstrated for this reaction in the gas phase \cite{Sha2012} that it is possible to obtain up to 60\% polarization at the sites of the two added protons.  This means that the overall polarization of the hydrogen nuclei on the substrate molecules is $\approx 0.6\times (2/8)=15$\%.  The theoretical maximum (for 100\% PHIP) would be 25\% overall polarization.  An improved choice may be ethylene (C$_2$H$_4$) to ethane (C$_2$H$_6$), which would yield a theoretical maximum of $\approx$33\%.  If one uses acetylene (C$_2$H$_2$) as the starting molecule, such a reaction will yield a mixture of ethylene and ethane.  The interesting product here would be polarized ethylene, with four hydrogen nuclei: two of which are polarized and two are unpolarized, yielding a potential theoretical maximum polarization of 50\%.  It may be possible to separate the desired product from the mixture and condense it to the liquid phase using a liquid nitrogen [boiling point (b.p.) 77 K] cold finger to get a high density of polarized hydrogen nuclei (b.p. of ethylene is 169.5 K).
We note that one way to obtain nuclear-spin order that can be converted into magnetic polarization in zero or low magnetic fields is to use a molecule that possesses an asymmetric $J$-coupling topology.
In the case of propane, it is sufficient that one of the carbons adjacent to the added protons be magnetic (e.g., $^{13}$C). Subsequent evolution of the spins of the protons and the magnetic nucleus under the influence of each other leads to net magnetization that can be generated after the application of a magnetic-field pulse.

Because selectively $^{13}$C-labeled molecules are expensive and experiments done in continuous-flow mode may rapidly consume molecules, it may be desirable to recycle the hydrogenated product (see Fig.\ \ref{Fig:Scheme}). To this end, we may envisage closing the circuit by adding a dehydrogenation step after the polarized target loses its polarization.  Dehydrogenation of propane and other alkanes is possible, for example, with the use of vanadium, molybdenum and tungsten oxides \cite{Che2000}, albeit catalysts will differ in their selectivity and the conditions of the dehydrogenation reaction (pressure, temperature) may be different.

\section{Optimization}
\label{Sec:Optimization}
In order to make an efficient polarized target, all aspects of the process must be optimized. The following is a partial list of parameters and issues that should be addressed.
\begin{itemize}
\item The choice of the substrate molecule. Our goal here is to have the highest fraction of polarized nucleons compared to the overall number of nuclides in the molecule in addition to having an overall high areal density of the target. The efficiency of polarization, polarization relaxation time (which we need as long as possible), the ability to flow and recycle the material are all important factors. Will the substrate be compressed gas or liquid, or, perhaps, a supercritical phase? Liquid state would be advantageous in terms of density and, typically, longer relaxation times than gas; however, elevated pressures and/or low operating temperatures may be required to suppress evaporation. An additional consideration is the temperature dependence of the hydrogenation reaction.
\item The choice of the catalyst that will ensure efficient hydrogenation on the time scale of spin relaxation. This is essential to have high spin polarization. Both homogenous and heterogeneous catalysts options are available.  Heterogeneous catalysts have the advantage that they can be packed into a reactor. The reacted product is free of catalyst contamination, which would be detrimental for this experiment.  There is also no need for a solvent in many heterogeneous reactions. The highly efficient catalyst for PHIP \cite{Sha2012} mentioned above meets these requirements.
\item The converter chemistry: If a heterogeneous catalyst is used, a catalytic converter (reactor) is needed. There are important issues that must be addressed when designing an appropriate reactor.  The reactor should be dense enough for efficient reaction, but not too dense as to hinder transport processes.  Efficient heat transport is required to avoid deterioration of the catalyst.  A short residence time is desirable in order to avoid depolarization of the product.
\item An efficient flow/shuttle/transport system is required to bring the reaction product to the target region.  Short transit times are needed to avoid depolarization. Sufficiently good control of the transport is needed to transport and subsequently immobilize the product.
\end{itemize}

\section{Conclusion and possible extensions}
A spin-polarized-target concept based on parahydrogen-induced polarization (PHIP) has been proposed. The PHIP targets may offer a number of advantages over the existing polarized targets, including fast polarization reversal and non-cryogenic operation at near-zero magnetic field. So far, no ``show stoppers'' are seen; however, specific technical details of the novel target still need to be worked out, which is partially a chemical-engineering task to be addressed in future work.

There are several interesting extensions of the basic concept that deserve further consideration:
\begin{itemize}
\item Nonhydrogenative (NH) PHIP \cite{Ada2009Science,Ada2009} expands the types of substrates that could be polarized, and, more importantly, eliminates the need for the dehydrogenation step. Note that a zero-field nonhydrogenative PHIP experiment has recently been performed \cite{The2012}. It is interesting to note in this context that in the work \cite{Glo2011}, where NH-PHIP of aminoacids and peptides was demonstrated, hyperpolarized HD molecules were also produced, which are an attractive polarized-target material.
\item The PASADENA/ALTADENA PHIP schemes (see, for example, the review \cite{Nat97}) that rely on relatively high magnetic field and removing the degeneracy between the two protons not by $J$-coupling, but by chemical-shift differences, may also be considered. The advantage here could be that protons are the only polarized nuclei (for example, no $^{13}$C nuclei are involved).
\item The possibility of doing the hydrogenation reaction in-situ should also be considered, which will likely involve homogeneous catalysis.  An advantage is that this would avoid sample transfer.
\item Photochemically induced reactions with parahydrogen should be considered. More generally, photochemically induced nuclear polarization (CIDNP \cite{Goe97}, without parahydrogen) that has been discussed in the context of polarized targets for over twenty years \cite{Pop89} should be revisited.
\item Finally, we briefly discuss PHIP polarized ethyl fluoride $^{12}$CH$_3$-$^{12}$CH$_2$-$^{19}$F. The initial pH$_2$ singlet polarization at zero field ``spreads'' over the whole molecule via the $J$-coupling network. The difference with the cases discussed in Sec.\ \ref{Sec:hydrogenation}, is that here, instead of $^{13}$C, it is $^{19}$F that breaks the symmetry between the protons, as revealed by application of the magnetic-field pulse. In this case, however, the oscillation frequency after the pulse is smaller due to the fact that the second-order H-F coupling is smaller than the first-order H-$^{13}$C coupling ($^2J_{HF} < ^1J_{HC}$). Now, the $^1$H chemical shift of $^{12}$CH$_3$ is different from that of $^{12}$CH$_2$-F (by a few ppm), so at moderate DC magnetic fields $\sim$1000 G, after the pulse, the magnetization will oscillate with the Larmor frequency of $^1$H and $^{19}$F (modulated by the $J$-coupling). This means a much faster reversal of the polarization (up to several MHz), but at a cost of applying a homogeneous magnetic field of $\sim$1000 G.
\end{itemize}

Finally, we note that the development of efficient nuclear polarization techniques may have important ramifications also outside of the polarized-target context, including the entire plethora of the applications of NMR and magnetic-resonance imaging, production of polarized proton beams, the studies of $\beta$-decay asymmetries, and in enhancement of nuclear-fusion reactions via nuclear polarization \cite{Kul82}.

\begin{acknowledgments}
This work was supported by the National Science Foundation under Grants CHE-0957655 and PHY-1068875. We are grateful to Max Zolotorev for stimulating discussions.
\end{acknowledgments}

\bibliographystyle{unsrt} 
\bibliography{NQR_NMR,PolarizedTargets}

\end{document}